\documentstyle[11pt,paspconf]{article} 

\begin{document} 

\title{$\lambda_{0}$ and $\Omega_{0}$ from Lensing Statistics and Other
Methods: Is There a Conflict?} 

\author{Phillip Helbig, Juan Francisco Macias-Perez, Althea Wilkinson,
Rod Davies} 

\affil{
University of Manchester, Nuffield Radio Astronomy Laboratories,
Jodrell Bank, Macclesfield, Cheshire SK11~9DL, UK
} 

\author{Ralf Quast}

\affil{
Hamburg Observatory, Gojenbergsweg 112, 21029 Hamburg, Germany
}

\begin{abstract}
We discuss the consistency of constraints in the
$\lambda_{0}$-$\Omega_{0}$ plane from gravitational lensing statistics,
the $m$-$z$ relation for type Ia supernovae and CMB anisotropies, based
on our own (published or unpublished) work and results from the
literature. 
\end{abstract}


Recently, several authors, (e.g.~Perlmutter et al.~1999, Riess et
al~1998, Lineweaver 1998) have made the claim that current observational
data provide evidence for a positive cosmological constant
$\lambda_{0}$.  This is based mainly on the $m$-$z$ relation for type~Ia
supernovae or on CMB anisotropies; although joint constraints from more
than one cosmological test also point in this direction (e.g.~Ostriker
\& Steinhardt 1995, Turner 1996, Bagla et al.~1996, Krauss 1998), taken
at face value, either the supernovae or CMB data \emph{alone} suggest
the presence of a positive cosmological constant. 

On the other hand, gravitational lensing statistics (e.g.~Kochanek 1996,
Falco et al.~1998) has often been seen as setting tight upper limits on
the cosmological constant, perhaps even to the point of being in
conflict with the `cosmic concordance' of, for example, a flat universe
with $\lambda_{0}\approx 0.7$ and $\Omega_{0}\approx 0.3$.  Based on the
present observational data, is there a conflict? 

We have compared contours in the two-dimensional space of the
$\lambda_{0}$-$\Omega_{0}$ plane, with no priors on $\lambda_{0}$ or
$\Omega_{0}$, from results based on our own work involving gravitational
lensing statistics and the CMB (Quast \& Helbig 1999, Helbig et
al.~1999, Helbig 1999, Macias-Perez et al.~1999, hereafter Papers~I--IV,
respectively) and from results from the $m$-$z$ relation for type Ia
supernovae, kindly made available by the Supernova Cosmology Project and
the High-Z Supernova Search Team (Perlmutter et al.~1999, Riess et
al.~1998).  

Are the results from lensing statistics and the $m$-$z$ relation for
type Ia supernovae consistent?  As the 90\% confidence contours from all
supernovae data sets overlap with that of the lensing statistics, and
even the 68\% confidence contours from two of three supernovae data sets
(one from Perlmutter et al.~(1999) and one, with two different methods
of analysis, from Riess et al.~(1999)) overlap with that of the lensing
statistics, the results from the two cosmological tests are consistent
and one is justified in calculating joint constraints by multiplying the
probability distributions of the individual tests.  Interestingly, they
are most consistent at small, but not too small, values of $\Omega_{0}$,
which is favoured on completely different grounds.  Joint constraints
from lensing statistics and the $m$-$z$ relation for type Ia supernovae
are discussed in detail in Paper~III. 

We have used the most recent CMB data available to do an analysis
similar to that of Lineweaver (1998) and compare the constraints from
the CMB to those of lensing statistics.  For more details, see Paper~IV.
The constraints from the CMB are much tighter than those from lensing
statistics or supernovae and a given confidence contour from the CMB is
always (almost) contained within the corresponding contours from the
other tests.  Thus, there is no inconsistency. 

The full poster can be obtained from\\
\texttt{http://multivac.jb.man.ac.uk:8000/ceres/papers/papers.html}\\
where one can also find related publications.  In the poster and related
papers, we show plots of constraints from various cosmological tests,
both from our own results as well as from the literature, in the same
area of parameter space and with the same scale, plotting scheme
etc., which makes comparison easy. 

\acknowledgments 
It is a pleasure to thank Saul Perlmutter, Brian Schmidt and Saurabh Jha
for helpful discussions and the Supernova Cosmology Project and the
High-Z Supernova Search Team for making their numerical results
available. We thank D.~Barbosa, G.~Hinshaw and C.~Lineweaver for helpful
discussions and M.~Zaldarriaga and U.~Seljak for making their CMBFAST
code publicly available. JFMP acknowledges the support of a PPARC
studentship.  Much of this poster is based directly or indirectly on the
efforts of the CLASS collaboration and those of the CERES EU-TMR
Network, coordinated by Ian Browne at Jodrell Bank, whose main purpose
is to make use of CLASS---among other things for studies of the
cosmological aspects of gravitational lensing.  This research was
supported in part by the European Commission, TMR Programme, Research
Network Contract ERBFMRXCT96-0034 `CERES'.

\end{document}